# Unidirectional output from a quantum-dot single-photon source hybrid integrated on silicon

Ryota Katsumi[1, 2, 3, 4], Yasutomo Ota[5, 6], Takeyoshi Tajiri[2, 7], Masahiro Kakuda[6], Satoshi Iwamoto[1, 2, 6], Hidefumi Akiyama[3] and Yasuhiko Arakawa[6]

[1] Research Center for Advanced Science and Technology, The University of Tokyo, 4-6-1 Komaba, Meguro-ku, Tokyo, Japan

[2] Institute of Industrial Science, The University of Tokyo, 4-6-1 Komaba, Meguro-ku, Tokyo, Japan

[3] Institute for Solid State Physics, The University of Tokyo, 5-1-5 Kashiwanoha, Kashiwa, Chiba, Japan

[4] Department of Electrical and Electronic Information Engineering, Toyohashi University of Technology, 1-1 Hibarigaoka, Tenpaku-cho, Toyohashi, Aichi, 441-8580 Japan

[5] Department of Applied Physics and Physico-Informatics, Faculty of Science and Technology, Keio University, 3-14-1 Hiyoshi, Kohoku-ku, Yokohama 223-8522, Japan

[6] Institute for Nano Quantum Information Electronics, The University of Tokyo, 4-6-1 Komaba, Meguro-ku, Tokyo, Japan

[7] Department of Computer and Network Engineering, University of Electro-Communications, 1-5-1 Chofugaoka, Chofu, Tokyo 182-8585, Japan

*Corresponding authors*: katsumi.ryota.ti@tut.ac.jp, ota@appi.keio.ac.jp

## Abstract

**We report a quantum-dot single-photon source (QD SPS) hybrid integrated on a silicon waveguide embedding a photonic crystal mirror, which reflects photons and enables efficient unidirectional output from the waveguide. The silicon waveguide is constituted of a subwavelength grating so as to maintain the high efficiency even under the presence of stacking misalignment accompanied by hybrid integration processes. Experimentally, we assembled the hybrid photonic structure by transfer printing, and demonstrated single-photon generation from a QD and its unidirectional output from the waveguide. These results point out a promising approach toward scalable integration of SPSs on silicon quantum photonics platforms.**

Integrated photonic circuits provide a versatile platform for photonic quantum information processing [1]. Among various material platforms, silicon photonics [2,3], which employs complementary-metal-oxide-semiconductor (CMOS) technology, is particularly attractive for constructing large-scale and highly-functional integrated quantum photonic circuits (IQPCs) [4,5]. Silicon quantum photonics (i.e., the utilization of silicon photonics for quantum applications) has shown great promise for quantum communication [6], quantum teleportation [7], and boson sampling [8]. To scale up silicon-based IQPCs with discrete variables, it is desirable to integrate deterministic single-photon sources (SPSs) on chip, rather than using probabilistic SPSs based on nonlinear optical processes.

An attractive approach to implement deterministic SPSs is the hybrid integration of quantum emitters on silicon [9], such as color centers in solids [10–12], defects in two-dimensional materials [13], and carbon nanotubes [14]. Especially, semiconductor quantum dots (QDs) are highly promising [15,16], since they have been proven to deterministically generate single photons with high purity and indistinguishability [17,18]. The emission wavelengths of QDs can be tuned to the telecommunication wavelength band [19–21], which is advantageous for exploiting well-developed silicon photonics components. To date, several techniques have been developed for the hybrid integration of QD SPSs onto photonic circuits made of silicon-based materials [22–28]. Among various hybrid integration approaches, transfer printing and micromanipulation offer unique routes of hybrid integration. These techniques are based on pick-and-place operation for integrating nanophotonic components, which can be conducted after completing the entire CMOS fabrication processes and thus are straightforwardly compatible with them [22,23,28–33]. Previously, we have demonstrated the hybrid integration of QD SPSs on a CMOS-processed silicon chip using transfer printing and confirmed optical and thermal coupling between the sources and photonic chips assembled by the pick-and-place technique [28].

In the previous demonstration, we investigated the device structure shown in Fig. 1(a). A QD-SPS with a photonic crystal nanocavity is transfer-printed on a glass-clad silicon waveguide. The cavity structure is symmetric and the waveguide is homogeneous along with the wave propagation direction so that the photon output in the waveguide is bidirectional, limiting the maximum possible efficiency of the source by half. This is highly detrimental to the scalable operation of IQPCs. It is possible to make the photon output unidirectional by employing directional couplers with adiabatic tapers [23,34] or spin-momentum locking via chiral light-matter coupling [35–37]. Another simple method is to embed a photonic mirror in the waveguide, which reflects photons and rectifies the flow of light in the waveguide [22,38]. However, the presence of the mirror could lead to destructive interference between photons bounced at the mirror and those were not [38]. To avoid such an unfavorable process, careful design and control of the source-mirror distance are necessary. This requirement poses a challenge when adopting the mirror for transfer-printed SPSs, since the integration technique inevitably accompanies non-negligible position misalignment.

In this work, we demonstrate unidirectional output from QD SPSs transfer-printed on a silicon waveguide terminated with a photonic crystal (PhC) mirror. We employed a subwavelength grating (SWG) waveguide, which reduces the effective refractive index of the silicon waveguide and relaxes the required positioning accuracy on the source-mirror distance. In design, we found that the proposed configuration can attain near-unity efficiencies of single- photon output while maintaining high robustness against positioning misalignment of the QD SPS. Experimentally, we fabricated the designed structure on a silicon photonic circuit and observed unidirectional output of single photons from an integrated single QD. The proposed approach offers a promising path to highly efficient and deterministic SPSs hybrid integrated on silicon-based IQPCs.

Figure 1(b) illustrates the investigated device structure in this work. A QD SPS is placed above a glass-clad silicon waveguide. The QD is embedded in a GaAs PhC nanobeam cavity (lattice constant = $a$, hole radius = $r$), which meditates the efficient coupling of QD emission into a cavity resonant mode through Purcell enhancement. The QD emission coupled to the cavity mode is then evanescently coupled to the underneath silicon waveguide [39]. This device structure allows for near-unity single-photon coupling from the QD into the silicon waveguide by controlling the vertical cavity-waveguide distance ($d$) and by establishing the phase matching between the cavity mode and waveguide mode [29]. The waveguide is designed to support a single transverse electric-like waveguide mode and is terminated at one end with a PhC mirror for realizing the unidirectional output of single photons. Here, we define +$x$ as the direction of the unidirectional photon output.

Photons trapped in the cavity mode can couple to the waveguide mode propagating in either +$x$ or −$x$ direction. Photons propagating in -$x$ direction are reflected by the PhC mirror (lattice constant $A = 0.95a$, hole radius $R = 0.28a$) and then start propagating in +$x$ direction, which interfere with photons directly coupled to +$x$ direction from the cavity. To efficiently reflect photons with suppressed light scattering, the sizes of the holes near the end of the PhC mirror are gradually modulated as shown in the inset of Fig. 1(b) ($R_1 = 0.94R$, $R_2 = 0.96R$, and $R_3 = 0.98R$). Here, we define the position of the mirror at the edge of the end air hole and that of the cavity at its center.

Photons that undergo a round trip between the cavity and the mirror with a distance of $L$ acquire a phase of $4\pi n_{eff} L/\lambda$ in addition to that accompanied by the reflection at the mirror, where $\lambda$ is the wavelength of the photon and $n_{eff}$ is the effective refractive index of the guided mode. As such, the constructive and destructive interferences sinusoidally occur with $L$. The elongation of the period of the oscillation is essential to design a waveguide-coupled SPS which is less insensitive to position misalignment. For this purpose, we diminish $n_{eff}$ by introducing a fish-bone type SWG waveguide schematically shown in the inset of Fig. 1(b). The SWG structure enables the control of $n_{eff}$ without changing the thickness or material of the waveguide, thereby being adaptable for standard CMOS processes for patterning photonic circuits. We set the thickness and width of the SWG waveguide to $1.14a$ and $0.5a$, respectively, which are the same with those in the PhC mirror region. The unit structure of the SWG is arranged with a pitch of $p = a/2$, which is so short compared to $\lambda/n_{eff}$ that the structure can be recognized as an effective medium and thus does not diffract the guided light. We tuned $n_{eff}$ by controlling the length $s$ and width $w$ of the narrow section of the SWG structure. In this section, we chose $s = 0.23a$ and $w = 0.2a$, which results in $n_{eff}$ of 1.7, which is 26% smaller than a normal square-shaped wire waveguide with the same thickness and width. Here, we assume the refractive index of silicon to be 3.5.

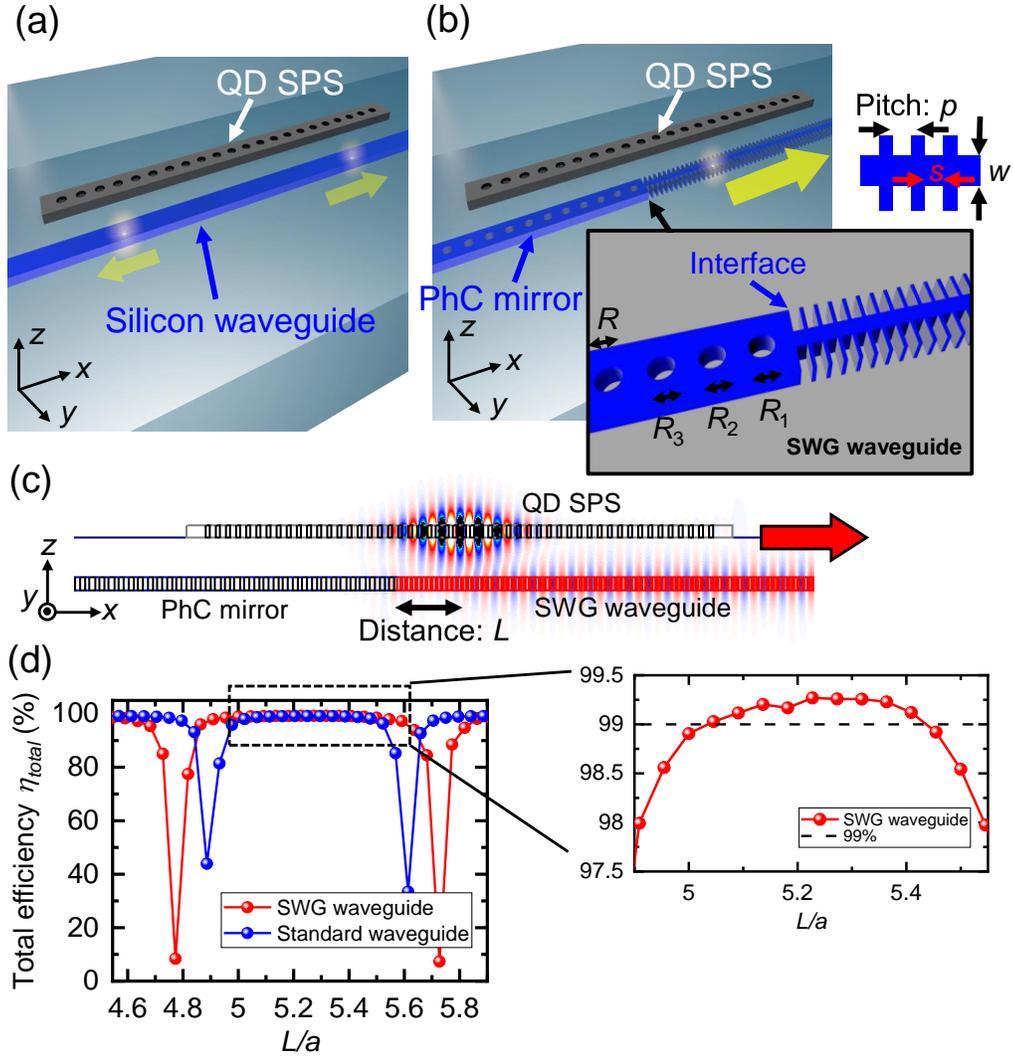

Fig. 1. (a) Schematic of a QD SPS structure coupling to a homogeneous wire waveguide. (b) Schematic of the investigated device structure. (c) Simulated field profile of the investigated cavity mode coupled to the underlying SWG waveguide terminated with the PhC mirror. (d) Simulated $\eta_{total}$s plotted as a function of $L$.

The GaAs PhC nanobeam cavity was designed to be phase-matched with the SWG waveguide. The thickness, width, and air hole radius were set to be $0.45a$, $1.02a$, and $0.23a$, respectively. We introduced a lattice spacing modulation around the cavity center to support a high $Q$-factor cavity mode. The modulation was applied under the same rule in our previous work [40] and resulted in a $Q$-factor of over $10^6$ and a mode volume of $0.57(\lambda/n)^3$ when placing the cavity above a plane glass layer without any waveguide underneath.

Then, we simulated the optical coupling of QD emission to the SWG waveguide using the finite difference time domain method. First, we considered a situation under the constructive interference condition that occurred at $L = 5.23a$ for the structure with an optimized glass-clad thickness of $d = 1.59a$. Figure 1(c) displays a simulated field profile of the investigated cavity mode coupled to the SWG waveguide terminated with the PhC mirror. In this simulation, we excited the fundamental cavity mode oscillating at a normalized frequency of $0.28a/\lambda$. It can be confirmed that the accumulated photons inside the cavity are coupled to the underneath SWG waveguide and unidirectionally propagate in the waveguide with negligible light scattering. We evaluated the total coupling efficiency of the QD radiation to the guided mode propagating to $+x$ direction ($\eta_{total}$) by monitoring the power distribution to the waveguide mode in comparison to the whole radiated power. In the simulator, we employed a dipole radiation source resonant to the cavity mode, which emulates a QD emitter. In this way, we can take into account the entire cascaded optical coupling processes from the QD to the waveguide mode propagating to $+x$ direction [29]. For the constructive condition, we found a very high $\eta_{total}$ of 99.3%. In this case, the cavity $Q$-factor exhibits a significant reduction to $Q = 1.8 \times 10^3$. This indicates that there was a strong optical coupling between the cavity and the waveguide mode. In contrast, for the case with the destructive interference at $L = 5.73a$, $\eta_{total}$ is calculated to be a significantly lower value of 7%. In this case, the $Q$-factor maintains a high value of $1.7\times10^5$, which suggests that the destructive interference hampers the optical coupling between the cavity and waveguide. From the two cases, it is concluded that the dominant factor determining the cavity $Q$-factor is its coupling to the waveguide in the current system.

Next, we in detail investigated the influence of the cavity position on $\eta_{total}$. We modified the cavity position along with the waveguide and simulated $\eta_{total}$ at each position. Figure 1(d) summarizes computed $\eta_{total}$ s plotted as a function of $L$. We observed a periodic change of $\eta_{total}$ with respect to $L$. For comparison, we also computed coupling efficiencies for a structure with a silicon wire waveguide (blue points). The plot was obtained after optimizing the cavity and clad thickness so as to maximize $\eta_{total}$, thus enabling a fair comparison among the results obtained from different structures. The period of the change is shorter for the case of the wire waveguide, as expected from its larger $n_{eff}$ than that of the SWG waveguide. This result suggests that the SWG waveguide is more robust with respect to the change of $L$ than the normal wire waveguide. As indicated in the inset of Fig. 1(d), $\eta_{total}$ is maintained over 99% even under the presence of a position derivation of up to $\pm 0.28a$. For $\lambda = 1,580$ nm ($a = 440$ nm), $\eta_{total}$ is maintained over 99% even with the position derivation of $\pm 90$ nm, which is well within the positioning accuracy of transfer printing ($\pm 50$ nm) [29]. We note that such a high $\eta_{total}$ can be maintained even when the cavity is misaligned to the direction normal to the wave propagation direction or rotated by a few degrees. These properties are highly beneficial for realizing hybrid-integrated QD SPSs using pick-and-place assembly.

We fabricated the designed device as follows. We first prepared InAs/GaAs QD SPSs and silicon SWG waveguides separately. We fabricated 1D PhC-based QD SPSs ($r$ = 78 nm, $a$ = 300 nm, width = 450 nm, slab thickness = 180 nm, resonant cavity wavelength ~ 1,170 nm) into a GaAs slab containing one layer of self-assembled InAs QDs emitting around a wavelength of 1,200 nm. We chose a proper set of device parameters to tune the device operation wavelength to that of the QDs. The 1D PhC nanobeam cavities were patterned through standard semiconductor nanofabrication processes such as electron beam lithography and dry and wet etching. In parallel, we fabricated SWG waveguides into a 220 nm-thick silicon slab through the above-mentioned semiconductor nanofabrication processes. Figure 2(a) displays a scanning electron microscope (SEM) image of a fabricated SWG silicon waveguide. The deformation of the SWG shape is predominantly due to the proximity effect during the electron beam lithography. The modified SWG structure was found to exhibit $n_{eff}$ of 1.92 in simulation, which is low enough for robust device integration using transfer printing. The left side of the SWG waveguide is terminated by a PhC mirror (lattice period = 300 nm, measured hole radius = 90 nm), resulting in a photonic bandgap with a wavelength range from 1,050 nm to 1,250 nm. We then formed an upper clad on the waveguide by a spin-on-glass process. The thickness of the glass clad above the waveguide (which equals $d$) was precisely controlled to be 450 nm by optimizing the amount of solvent in the liquid glass material and the speed of spin coating [29]. To directly extract waveguide-coupled QD emission into free space, both ends of the waveguide were terminated with grating output ports.

Next, we utilized transfer printing to integrate fabricated QD SPSs onto SWG silicon waveguides. Figure 2(b) shows a schematic describing the transfer printing process. One of the prepared QD SPSs was picked up and then placed on the waveguide by a polydimethylsiloxane (PDMS) rubber stamp. The pick-up process was done by quickly pealing the PDMS stamp off. Solely the QD SPS was left on the waveguide by releasing the PDMS stamp by a slow peel motion. The QD SPS is firmly bonded on the chip mainly through van der Waals force. These pick-and-place processes are performed with a homemade transfer-printing apparatus composed of precision motion-controlled stages operating under an optical microscope [29]. Using the marker patterned on the waveguide layer, we performed transfer printing to place the cavity 2.16 μm from the PhC mirror, which was calculated to attain the constructive interference in the SWG waveguide. Figure 2(c) displays a visible microscope image of a completed device. Precise position alignment between the top nanobeam cavity and underlying waveguide can be seen. The position derivation between the nanobeam and waveguide is deduced to be < 50 nm regarding $y$ direction, which is routinely possible with our transfer-printing system [29].

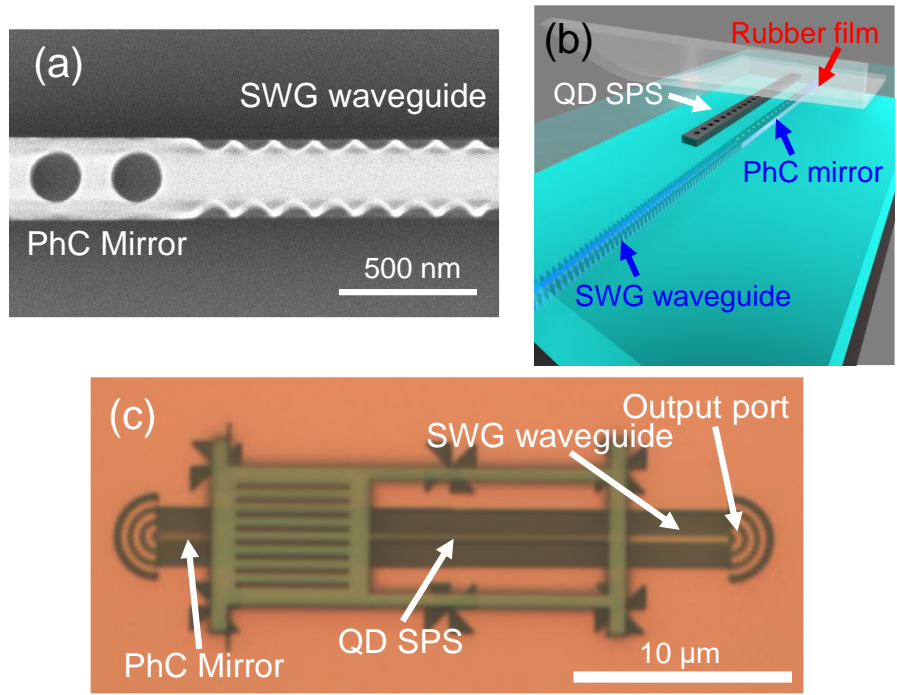

Fig. 2. (a) SEM image of a fabricated SWG waveguide terminated with a PhC mirror. (b) Schematic illustrating the procedure of transfer printing. (c) Visible microscope image of a completed device.

We characterized the fabricated device using low-temperature micro-photoluminescence (µPL) measurements. The devices were mounted in a liquid helium flow cryostat and kept at cryogenic temperatures. An objective lens with ×50 magnification and 0.65 numerical aperture was used for imaging the devices, focusing pump laser on the device, and collecting PL signals. The collected PL signals were analyzed using a spectroscope and an InGaAs camera. The bottom panel of Fig. 3(a) shows a PL image taken at 17 K with spectral band-pass filtering around the fundamental cavity mode (1,150 ± 20 nm). We obtained this image using a continuous-wave diode laser oscillating at 785 nm for pumping the cavity center with a strong pump power of 180 µW. Bright out-coupling of light was observed only from the right output port (red dotted circle), which indicates the unidirectional output from the QD light source into the underneath SWG silicon waveguide. Then, we reduced pump power to 2 µW and measured a PL spectrum via the right output port, as shown by the red curve in Fig. 3(b). The PL spectrum exhibits a peak of the cavity mode emission (1,144 nm) and those from a cavity-coupled QD. We label the prominent peak as QD1. The black solid line displays the cavity mode's peak, which was deduced through a Lorentzian peak fitting to a spectrum taken under strong pumping. In contrast, the spectrum measured above the cavity center (green curve) does not include any cavity peak. This indicates that the cavity leakage into free space is largely suppressed since the cavity emission is predominantly funneled into the SWG waveguide. We deduced an experimental $Q$-factor to be $Q_{exp}$ = 2,300. Meanwhile, the average $Q$-factor of the nanobeam cavities when placed on plane glass was found to be $Q_{ave}$ = 12,500 (average of ten samples). From these values and the assumption that the observed reduction of $Q_{exp}$ from $Q_{ave}$ is predominantly due to the cavity photon leakage into the underneath SWG waveguide, cavity-waveguide coupling efficiency was estimated to be 84% [29]. We note that similar reductions of cavity $Q$-factors to 1,500 ~ 4,000 were also observed in the other 7 samples we fabricated.

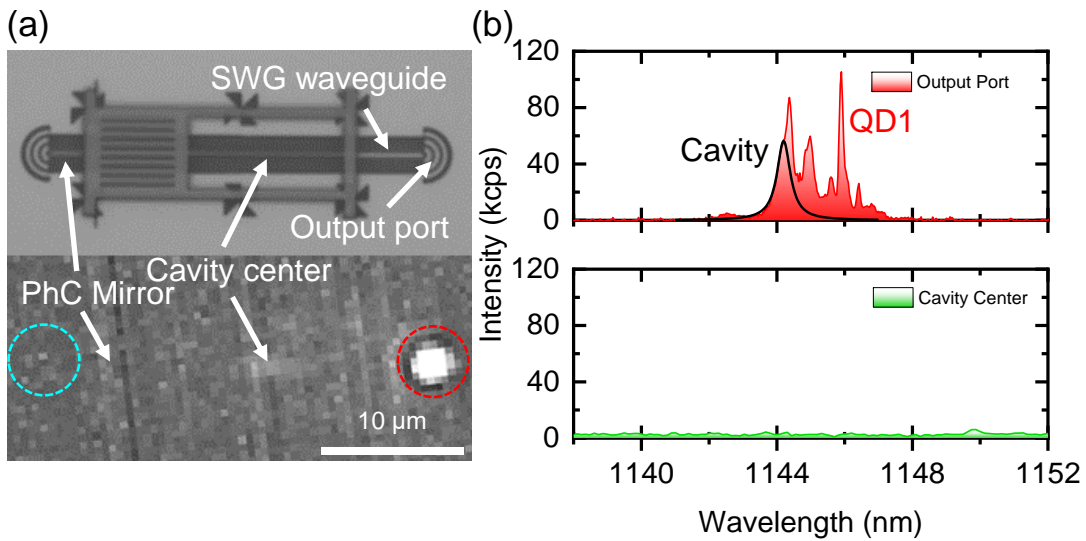

Fig. 3. (a) (Top panel) Visible microscope image of the fabricated device. (Bottom panel) PL image taken under the irradiation of a pump laser onto the cavity center. (b) PL spectra measured at the right output port (top panel) and above the cavity center (bottom panel).

To investigate the influence of the photon interference on the cavity-SWG waveguide coupling, we studied evolutions of cavity $Q$-factor when largely varying sample temperature, which effectively changes the cavity-mirror distance through the increase of the refractive indices of GaAs and Si. Figure 4(a) summarizes the measured $Q$-factors for the fabricated devices as a function of sample temperature. Here, Sample 1 is the same device discussed using Fig. 3(a) and is a representative of the case exhibiting the constructive interference at the lowest temperature as shown in Fig. 4(b). Sample 2 was fabricated by slightly shifting the cavity $x$ position by 70 nm compared to Sample 1 for inducing the destructive interference around the lowest temperature. Sample 1 shows an increase of cavity $Q$-factor when raising the sample temperature, which indicates that the device sustained the constructive interference at the lowest temperature as expected and was deviated from the constructive interference condition at higher temperatures. On the other hand, Sample 2 exhibits a much higher $Q$-factor of ~5,000 than that of Sample 1 at the lowest temperature and shows a decrease of $Q$-factors with elevating temperature. This suggests that Sample 2 was originally around the destructive interference condition and was tuned towards the constructive condition with increasing temperature. To further analyze the observed behaviors, we computed $Q$-factors by taking into account the temperature dependences of refractive indices of the constituent materials. We reproduced the device arrangements corresponding to Sample 1 and 2 in the simulator and emulated temperature rise in it. Figure 4(c) summarizes the simulated $Q$-factors of the cavities as a function of temperature. For Sample 1 (2), an increase (decrease) of $Q$-factors was well reproduced, the tendency of which agrees well with the experimental result. These results confirm the validity of our discussion provided above and that Sample 1 indeed sustained the constructive interference condition at the low temperature that we performed the PL measurements.

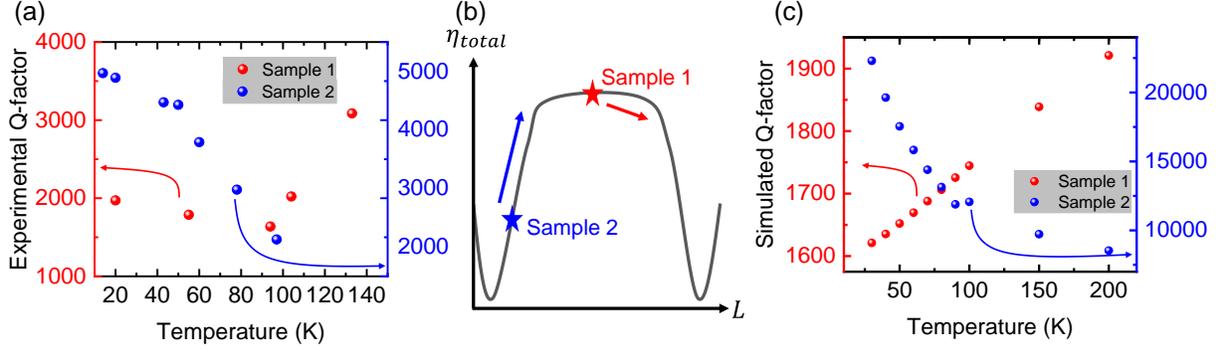

Fig. 4. (a) Experimental $Q$-factors of the cavities of Sample 1(red) and 2(blue) as a function of sample temperature. (b) Schematic illustration of the total coupling efficiency $\eta_{total}$ as a function of $L$. The red and blue stars show the situations of Sample 1 and 2, respectively. (c) Simulated $Q$-factors of the cavities corresponding to Sample 1 and 2 plotted as a function of sample temperature.

Next, we performed time-domain characterization of QD1 peak found in Sample 1. The cavity center was pumped using a pulsed laser (wavelength = 785 nm, averaged pump power = 200 nW, repetition rate = 80 MHz). We measured QD1 peak through the right output port at 6 K. The QD1 emission was spectrally isolated with a spectrometer and detected by a superconducting single-photon detector, resulting in a time-resolved PL spectrum as shown by the red curve in Fig. 5(a). We fit the measured data for QD 1 with a double exponential decay curve convolved with a function that reflects the system time response. Through this process, the spontaneous emission rate of QD 1 was deduced to be 3.8 GHz. We found that the emission rate of QD 1 becomes slower as the peak is detuned from the cavity resonance (gray curve), confirming that the emission of QD1 is enhanced by the Purcell effect. Meanwhile, the average radiative decay rate of QDs in unpatterned regions is only 0.9 GHz. Based on these experimental results after taking into account the photonic bandgap effect of the 1D nanobeam PhC structure [41], we deduced a QD-cavity coupling efficiency of ~88%. Combined with the estimated cavity-waveguide coupling efficiency in the previous section, we can estimate an overall QD-to-waveguide coupling efficiency of ~74% in this particular sample.

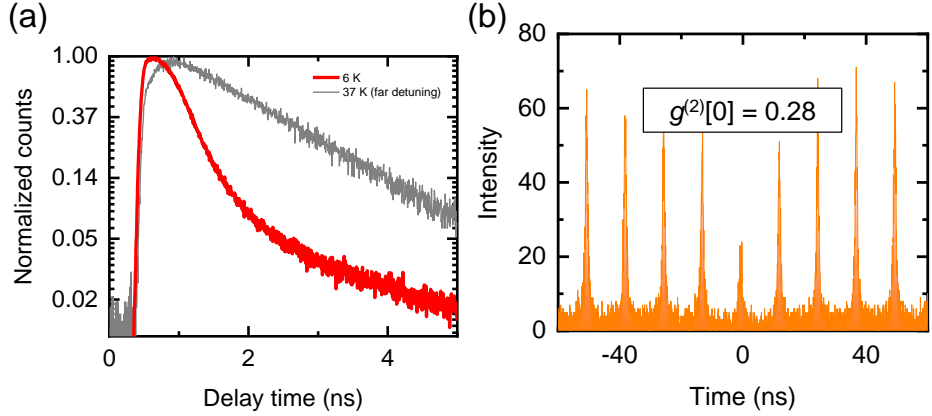

Fig. 5. (a) Time-resolved PL spectra measured when the QD is tuned near (red) and far (gray) from the cavity resonance. (b) Measured second-order coherence function.

Finally, to confirm the nonclassical nature of the QD emission, we performed Hanbury-Brown and Twiss correlation measurements for QD 1 peak. For this purpose, we added a 50:50 fiber beam splitter and a single-photon detector to the photon detection setup. Figure 5(b) shows an auto-correlation function $g^{(2)}[t]$ for QD1 peak measured under an average pump power of 200 nW. A clear anti-bunching with $g^{(2)}[0] = 0.28$ was observed, verifying the single-photon emission of the QD on silicon. We consider that the non-zero value of $g^{(2)}[0]$ is probably stemming from the intrusion of cavity background emission supplied by other off-resonant QDs inside the cavity.

In summary, we demonstrated a transfer-printed QD SPS capable of unidirectional single photon output in an underlying SWG silicon waveguide terminated with a PhC mirror. We designed the SPS structure to support a QD-to-waveguide coupling efficiency of > 99% by controlling the light interference in the waveguide. We introduced a SWG structure for the waveguide so as to make the structure robust to position misalignment accompanied by transfer printing. Experimentally, we succeeded in single-photon generation from a hybrid-integrated QD on silicon and its efficient coupling into an underlying SWG waveguide. The combination of the device design in this work and device assembly by transfer printing will provide a powerful route to the scalable implementation of efficient QD SPSs on highly functional silicon IQPCs.

**Acknowledgments:** This work was supported by JSPS KAKENHI Grant-in-Aid for Specially Promoted Research (15H05700), KAKENHI (18J21667, 19K05300), and JST PRESTO (JPMJPR1863), and based on results obtained from a project, JPNP13004, commissioned by the New Energy and Industrial Technology Development Organization (NEDO).